# Ion Acceleration by the Radiation Pressure of a Slow Electromagnetic Wave

S. V. Bulanov[a,b], T. Zh. Esirkepov[a], M. Kando[a], F. Pegoraro[c],

S. S. Bulanov[d], C. G. R. Geddes[e], C. Schroeder[e], E. Esarey[e], W. Leemans[e]

[a]QuBS, Japan Atomic Energy Agency, Kizugawa, Kyoto, 619-0215, Japan

[b]A. M. Prokhorov Institute of General Physics RAS, Moscow, 119991, Russia

[c]Physical Department, University of Pisa, Pisa, 56127, Italy

[d]University of California, Berkeley, CA 94720, USA

[e]Lawrence Berkeley National Laboratory, Berkeley, California, 94720, USA

**Abstract**

When ions are accelerated by the radiation pressure of a laser pulse, their velocity can not exceed the pulse group velocity which can be considerably smaller than the speed of light in vacuum. This is demonstrated in two cases corresponding to a thin foil target irradiated by high intensity laser light and to the hole boring produced in an extended plasma by the laser pulse that is accompanied by the formation of a collisionless shock wave. It is found that the beams of ions accelerated at the collisionless shock wave front are unstable against Buneman-like and Weibel-like instabilities which result in the broadening of the ion energy spectrum.

**1. Introduction**

Progress in the development of high power laser technology makes the idea of transferring momentum from light to macroscopic objects feasible. This idea goes back to Lebedev and Eddington [1]. In the mid 1950's ion acceleration by strong electromagnetic wave was suggested by V. I. Veksler [2]. He proposed to use the radiation pressure of a high intensity electromagnetic wave acting on an electron cloud that drags a portion of the ions by means of a collective electric field. The radiation pressure of a super-intense electromagnetic pulse on a thin plasma slab was studied in Ref. [3] as an acceleration mechanism able to provide ultrarelativistic ion beams. In this Radiation Pressure Dominant Acceleration (RPDA) regime the ions move forward with almost the same velocity as the electrons and thus have a kinetic energy well above that of the electrons. The laser interaction with an accelerated plasma slab can be considered as the light reflection from a receding mirror moving with relativistic velocity, when the energy and momentum of the reflected wave is much smaller than that of the incident wave. For this reason the process of laser energy transfer to the ions is highly efficient, with the ion energy per nucleon being proportional in the ultrarelativistic limit to the electromagnetic pulse energy.

An analytical description of the dynamics of a charged particle under the radiation pressure can be found in Ref. [4] (chapter 9, problem 6). There is an analogy between the RPDA mechanism and the "Light Sail" scheme for the spacecraft propulsion [5]. This scheme, which uses the transfer of momentum from the photons to the light-sail, has been proposed by F. A. Zander in 1924 [6]. The use of lasers for propelling a sailcraft over interstellar distances has been considered in Ref. [7] (for details and further discussions see Ref. [8]).

Laser-driven fast ions attract attention due to their broad range of applications ranging from medicine to high energy physics [9]. Since at present the RPDA is considered to be the main ion acceleration mechanism for the next generations of high power lasers, it has been studied by a number of scientific groups. In Refs. [10, 11] the stability of the accelerated foil has been analyzed. Refs. [12] are devoted to extending the RPDA range operation towards lower electromagnetic wave intensities. The radiation friction effects were incorporated into the RPDA model [13, 14]. An indication of the effect of the radiation pressure on bulk target ions is obtained in experimental studies of plasma jets ejected from the rear side of thin solid targets irradiated by ultraintense laser pulses [15, 16].

It was shown in Refs. [3, 17] that a foil interacting with a laser pulse becomes deformed and changes into a cocoon, which, in turn, traps the electromagnetic wave, thus allowing ion acceleration over a distance larger than the Rayleigh length. Cocoon-like structures lead to the enhancement of the energy of the accelerated ions[18]. Moreover it was shown in Ref. [18] that the use of targets expanding transversally increases the accelerated ion energy. The transverse expansion of the accelerated ion shell can be provided by the action of the ponderomotive force of a laser pulse with a finite waist. It can also occur as a result of the instability described in Ref. [10]. There are also regimes of laser ion acceleration, where the ion energy is enhanced by the target structuring [19].

On the other hand, when the laser pulse is confined inside cocoon-like structures, or inside a channel (it can be a self-focusing channel or a guiding structure inside a capillary or a cavity [20]), as well as in an underdense plasma, its group velocity is smaller than the speed of light in vacuum. Below we shall refer to such a wave as a "slow electromagnetic wave". Previously the RPDA regime of the laser ion acceleration was studied within the framework of a model that assumed the laser group velocity to be equal to the speed of light in vacuum [3, 10 - 12, 14, 17 - 19]. Kinematic considerations, similar to those that have been used in Ref. [10], show that in the case of ion acceleration by a slow electromagnetic wave the ion velocity cannot exceed the laser group velocity because the photons can no longer reach the receding mirror. Below we formulate the basic theory of the ion acceleration in this regime. We consider two cases corresponding to a thin foil target irradiated by a high intensity laser pulse and to the hole boring produced in the extended plasma by the laser pulse, accompanied by the formation of a collisionless shock wave corresponding to the regimes considered in Refs. [21–25]. Contrary to the case of a thin foil target, the ions accelerated at the shock wave front propagate further through relatively high density plasma. In such a configuration Buneman-like and Weibel-like instabilities develop [26] which can lead to the broadening of the ion energy spectrum. We describe these instabilities in the case of counter-propagating ion beams when the perturbations are driven by the shock wave front.

## 2. Dynamics of a Plasma Slab Accelerated by a Slow Electromagnetic Wave

The nonlinear dynamics of a laser accelerated foil is described within the framework of the thin shell approximation first formulated by E. Ott [27] and further generalized in Refs. [10, 28]. In the interaction of an electromagnetic wave with a thin foil, the latter is modelled as an ideally reflecting mirror. The equations of motion of the surface element of an ideally reflecting mirror in the laboratory frame of reference can be written in the form

$$\frac{d\bm{p}_\alpha}{dt} = \frac{\mathcal{P}}{\sigma}\bm{\nu}, \qquad (1)$$

where $\bm{p}_\alpha$, $\mathcal{P}$, $\bm{\nu}$, and $\sigma$ are the momentum, light pressure, unit vector normal to the shell surface element, and surface density, $\sigma = nl$, respectively. Here $n$ and $l$ are the plasma density and shell thickness. In order to describe how the shell shape and position change with time we introduce the Lagrange coordinates $\eta$ and $\zeta$ playing the role of the markers of a shell surface element. A surface element $\delta s$ carries $N = \sigma \delta s$ particles, where $N$ is constant in time and equal to $N = \sigma_0 \, d\eta \, d\zeta$. The equations of the shell element motion are [18]

$$\sigma_0 \partial_t p_{\alpha,i} = \mathcal{P} \varepsilon_{ijk} \partial_\eta x_{\alpha,j} \partial_\zeta x_{\alpha,k}, \qquad (2)$$

$$\partial_t x_{\alpha,i} = c \frac{p_{\alpha,i}}{\sqrt{m_\alpha^2 c^2 + p_{\alpha,k} p_{\alpha,k}}}. \qquad (3)$$

Here $m_\alpha$ is the ion mass, $\sigma_0 = n_0 l_0$ is the surface density at $t = 0$, $\varepsilon_{ijk}$ is the fully antisymmetric unit tensor, $i = 1, 2, 3$, and summation over repeated indices is assumed.

Assuming that the shell moves along the $x$ axis, i.e., that the initial conditions correspond to a plane slab homogeneous along the $y$ and $z$ axes, and setting $y_\alpha^{(0)} = \eta$ and $z_\alpha^{(0)} = \zeta$, we rewrite the equation for the $x$ component of the momentum (2) as $dp_{\alpha,x}^{(0)}/dt = \mathcal{P}/\sigma_0$, where $p_x^{(0)}$ depends on time only. The electric field at the moving shell (i.e., at $x = x(t)$) depends on time as $E(t-x(t)/v_g)$, where $v_g$ is the wave group velocity. The function $x(t)$ in this equation should be found from Eq. (3).

The force acting on the shell is expressed in terms of the flux of the electromagnetic wave momentum, which is proportional to the Poynting vector, $\mathbf{S} = c\,\mathbf{E}\times\mathbf{B}/4\pi$. Considering a circularly polarized electromagnetic wave with frequency $\omega$ and wave number $k$ propagating along the $x$ axis given by the vector potential

$$\mathbf{A} = A_0\left[\cos(\omega t - kx)\mathbf{e}_y + \sin(\omega t - kx)\mathbf{e}_z\right], \qquad (4)$$

and calculating the electric, $\mathbf{E} = -\partial_t \mathbf{A}/c$, and magnetic, $\mathbf{B} = \nabla\times\mathbf{A}$, fields, we find the Poynting vector

$$\mathbf{S} = c\,\omega k A_0^2 \mathbf{e}_x, \qquad (5)$$

proportional to the product of wave frequency and wave number, $\omega$ and $k$. In the boosted frame of reference moving with velocity $\beta$ we obtain for the product of wave frequency and wave number,

$$\bar{\omega}\bar{k} = \omega k\frac{1+\beta^2}{1-\beta^2} - (\omega^2+k^2)\frac{\beta}{1-\beta^2} = \omega^2\frac{(\beta_g-\beta)(1-\beta_g\beta)}{1-\beta^2}. \qquad (6)$$

In order to obtain the last term in the r.h.s. of Eq. (6) we used the relationship between the frequency, wave number and group velocity of an electromagnetic wave: $v_g = c\beta_g = kc^2/\omega$. Here and below a bar is used for variables in the boosted frame of reference.

The force acting on a thin foil is given by

$$F = (1+|\rho|^2 - |\tau|^2)S, \qquad (7)$$

where $|\rho|^2$ and $|\tau|^2$ are the light reflection and transmission coefficients, respectively, related to each other as

$$|\rho|^2 + |\tau|^2 + |\alpha|^2 = 1. \qquad (8)$$

Here $|\alpha|^2$ is the absorption coefficient. Using these relationships we obtain that the equation for the $x$ component of momentum (2) can be written in the form

$$\frac{1}{(1-\beta_\alpha^2)^{3/2}}\frac{d\beta_\alpha}{dt} = \left(\frac{K\,E^2}{4\pi\sigma_0 m_\alpha c}\right)\frac{(\beta_g - \beta_\alpha)(1-\beta_g\beta_\alpha)}{(1-\beta_\alpha^2)}, \qquad (9)$$

where $K = 2|\rho|^2 + |\alpha|^2$. In Eq. (9) $\beta_\alpha = p_{\alpha,x}/(m_\alpha^2 c^2 + p_{\alpha,x}^2)^{1/2}$ is the shell normalized velocity, $\beta_g = v_g/c$, and $E^2 = (\omega A/c)^2$. As we see, the radiation pressure vanishes at $\beta_\alpha = \beta_g$. In addition, in the case of complete absorption of the light, which corresponds to $|\rho|^2 = 0$ and $|\alpha|^2 = 1$, the radiation pressure at the foil is two times smaller than in the case of an opaque foil with no absorption. A detailed analysis of the foil opaqueness effect on the ion acceleration by the radiation pressure is presented in Ref. [29].

For a finite-length electromagnetic pulse, the electric field at the moving shell (i.e., at $x = x(t)$), depends on time as $E = E(t - x(t)/c\beta_g)$. Introducing the phase of the laser pulse at the shell, $\psi = \omega_0(t - x(t)/c\beta_g)$, as the new independent variable, we rewrite Eq. (9) in the form

$$\frac{dp_\alpha}{dw} = \frac{\beta_g \gamma_\alpha - p_\alpha}{\gamma_\alpha}, \tag{10}$$

where $w(\psi)$ is proportional to the laser pulse fluence,

$$w(\psi) = \int_0^\psi \frac{K E^2(\psi')}{4\pi\sigma_0 m_\alpha c} d\psi'. \tag{11}$$

Integrating Eq. (10) we obtain for the normalized velocity as a function of $w(\psi)$,

$$\beta_\alpha = \frac{(1+\beta_g)\tanh^2\left(w\sqrt{\frac{1+\beta_g}{1-\beta_g}} + \tanh^{-1}\sqrt{\frac{1-\beta_g}{1+\beta_g}}\right) - (1-\beta_g)}{(1+\beta_g)\tanh^2\left(w\sqrt{\frac{1+\beta_g}{1-\beta_g}} + \tanh^{-1}\sqrt{\frac{1-\beta_g}{1+\beta_g}}\right) + (1-\beta_g)}. \tag{12}$$

In the case of a long enough laser pulse in the limit $w(\psi) \to \infty$, Eq. (12) yields $\beta_\alpha \to \beta_g$.

For an electromagnetic wave with constant amplitude, $E = E_0 = $ constant, by integrating Eq. (9) we obtain the implicit dependence of the foil velocity on time

$$\ln\left|\frac{\beta_g\left[1-\beta_g\beta+\sqrt{(1-\beta_g^2)(1-\beta^2)}\right]\left[1-\beta_g\beta+(1-\beta)\sqrt{1-\beta_g^2}\right]}{(\beta_g-\beta)\left[1+\sqrt{1-\beta_g^2}\right]\left[1-\beta_g\beta+(1+\beta)\sqrt{1-\beta_g^2}\right]}\right| = \frac{(1-\beta_g^2)^{3/2}}{\beta_g}\frac{t}{\tau_{1/3}}, \tag{13}$$

where the characteristic time scale is given by

$$\tau_{1/3} = \frac{2\pi\sigma_0 m_\alpha c}{K E_0^2}, \tag{14}$$

and the integration constant in Eq. (13) is chosen so as to fulfil the initial condition $\beta_\alpha|_{t=0} = 0$. We see that the shell velocity cannot exceed the wave group velocity approaching it at $t \to \infty$ as

$$\beta_\alpha \approx \beta_g - \exp\left[-\frac{(1-\beta_g^2)^{3/2}}{\beta_g}\frac{t}{\tau_{1/3}}\right]. \tag{15}$$

For the description of the reflection at the relativistic mirror of an electromagnetic wave with group (and phase-) velocity not equal to the speed of light in vacuum see Ref. [30].

We note that, since in the limit $\beta_\alpha \to \beta_g$ the laser pulse is more and more detached from the foil, the foil becomes stable with respect to the transverse perturbations that lead to the foil break up into the density bubbles and lamps [10].

In order to recover the case when the wave group velocity is equal to the speed of light in vacuum, i.e. $v_g \to c$, we rewrite Eq. (9) in the form

$$\int \frac{\left(e^{3u}+e^u\right)du}{\left(1+\beta_g\right)^2-\left(1-\beta_g\right)^2 e^{4u}} = \frac{t}{2\tau_{1/3}}, \tag{16}$$

where $u = \tanh^{-1}\beta_\alpha$ is the rapidity. As we can easily see, the limit $\beta_g \to 1$, i.e. $e^{2u} \ll \left(1+\beta_g\right)/\left(1-\beta_g\right)$, corresponds to the case considered in Refs. [3, 10] which gives $p_\alpha \approx \left(3t/4\tau_{1/3}\right)^{1/3}$ for the ion momentum. The singularity in the denominator of the integrand in Eq. (16), $e^{2u} \to \left(1+\beta_g\right)/\left(1-\beta_g\right)$, corresponds to the dependence given by Eq. (16).

## 3. Hole Boring by Radiation Pressure

We consider the laser pulse interaction with an extended plasma target, which can formally be considered to be semi-infinite, under the conditions when the laser radiation pressure drives the plasma ahead, acting as a piston. This regime is called 'hole-boring' [13, 31].

In the case of the interaction of a constant amplitude laser pulse with a homogeneous plasma, a stationary solution exists which describes the plasma-vacuum interface (it is also called "collisionless shock wave" and has been studied in Refs. [21 - 25]) moving with constant velocity $(d\beta/dt=0)$. In the frame of reference moving with the velocity $c\beta$ the surface position is determined by the balance between the flux of electromagnetic wave momentum (see Eqs. (5 - 7)) and the ion momentum flux. This yields for $\beta$

$$\beta = \frac{\sqrt{4\beta_g \mathrm{B}_E^2 + \left(1-\beta_g^2\right)^2 \mathrm{B}_E^4} - \mathrm{B}_E^2\left(1+\beta_g^2\right)}{2\left(1-\beta_g \mathrm{B}_E^2\right)}, \tag{17}$$

with $\mathrm{B}_E = \left(E^2/2\pi n_0 m_\alpha c^2\right)^{1/2}$. The electric field, $E$, is taken in the laboratory frame of reference. Here we assume that $|\rho|^2 = 1$ and $|\alpha|^2 = 0$, i.e. $K = 2$, and take into account the fact that in the rest frame of the shock wave the plasma density is equal to $n = n_0/\sqrt{1-\beta^2}$. It follows from Eq. (14) that the shock wave velocity cannot exceed the group velocity of the electromagnetic wave, $\beta \leq \beta_g$. In the case $\beta_g = 1$ we recover the result obtained in Ref. [13], which gives $\beta = B_E / (1 + B_E)$.

In the boosted frame of reference, $\mathcal{M}$, moving with velocity $\beta c$, the ions bounce at the laser pulse front. Their momentum changes from $-m_\alpha c \beta / \sqrt{1-\beta^2}$ to $m_\alpha c \beta / \sqrt{1-\beta^2}$, as illustrated in Fig. 1. In the laboratory frame of reference, $\mathcal{L}$, the momentum and the energy of the fast ions are equal to

$$p_\alpha = m_\alpha c \frac{2\beta}{1-\beta^2} \quad \text{and} \quad \mathcal{E}_\alpha = m_\alpha c^2 \left( \frac{1+\beta^2}{1-\beta^2} \right), \tag{18}$$

i.e.

$$p_\alpha = m_\alpha c \frac{2\left(1-\beta_g B_E^2\right)\left[\sqrt{4\beta_g B_E^2 + \left(1-\beta_g^2\right)^2 B_E^4} - \left(1+\beta_g^2\right) B_E^2\right]}{4\left(1-\beta_g B_E^2\right)^2 - \left[\sqrt{4\beta_g B_E^2 + \left(1-\beta_g^2\right)^2 B_E^4} - \left(1+\beta_g^2\right) B_E^2\right]^2} \tag{19}$$

and

$$\mathcal{E}_\alpha = m_\alpha c^2 \frac{4\left(1-\beta_g B_E^2\right)^2 + \left[\sqrt{4\beta_g B_E^2 + \left(1-\beta_g^2\right)^2 B_E^4} - \left(1+\beta_g^2\right) B_E^2\right]^2}{4\left(1-\beta_g B_E^2\right)^2 - \left[\sqrt{4\beta_g B_E^2 + \left(1-\beta_g^2\right)^2 B_E^4} - \left(1+\beta_g^2\right) B_E^2\right]^2}. \tag{20}$$

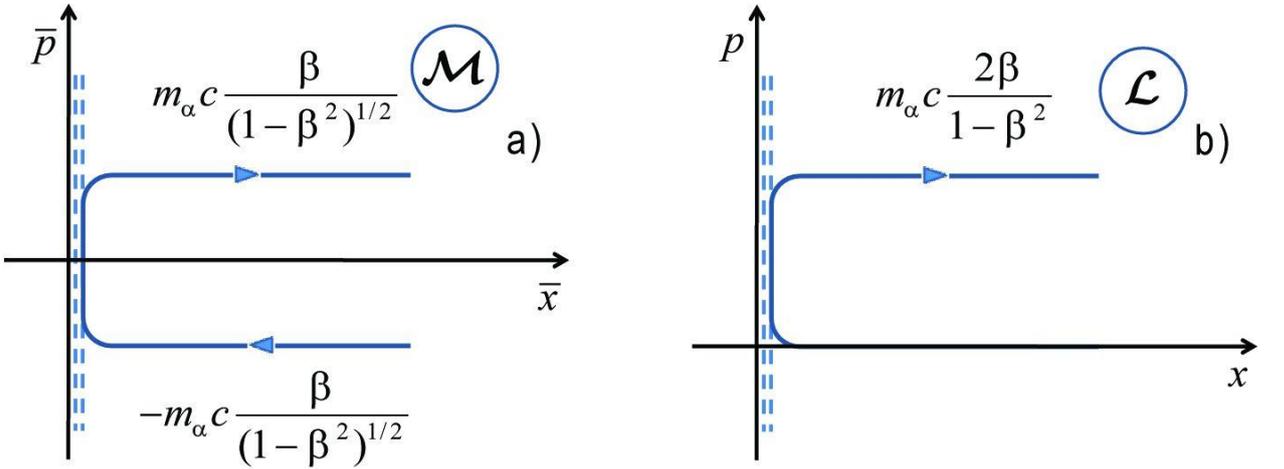

Fig. 1. Phase plane of the ions accelerated at the front of the laser pulse interacting with a homogeneous plasma. a) In the boosted frame of reference, $\mathcal{M}$, the ion momentum changes from $-m_\alpha c \beta / \sqrt{1-\beta^2}$ to $m_\alpha c \beta / \sqrt{1-\beta^2}$. b) In the laboratory frame of reference, $\mathcal{L}$, the ion momentum changes from zero to $2 m_\alpha c \beta / (1-\beta^2)$.

In Fig. 2 we show the ion kinetic energy, $\gamma_\alpha - 1$, as a function of the laser pulse amplitude, $B_E$, for $\beta_g = 1$ and $\beta_g < 1$. In the case $\beta_g = 1$ the kinetic energy of the ions accelerated at the collisionless shock front,

$$\mathcal{E}_\alpha = m_\alpha c^2 (\gamma_\alpha - 1) = 2 m_\alpha c^2 \frac{B_E^2}{1 + 2 B_E} \quad (21)$$

asymptotically as $B_E \to \infty$ is proportional to $B_E$. In the limit $B_E \ll 1$ we have for the energy scaling of the fast ions

$$\mathcal{E}_\alpha = \left( \frac{I}{2.5 \times 10^{21} W/cm^2} \right) \left( \frac{10^{21} cm^{-3}}{n_0} \right) GeV. \quad (22)$$

In the case $B_E \gg 1$ we obtain

$$\mathcal{E}_\alpha = \left( \frac{m_\alpha}{m_p} \right)^{1/2} \left( \frac{I}{5 \times 10^{21} W/cm^2} \right)^{1/2} \left( \frac{10^{21} cm^{-3}}{n_0} \right)^{1/2} GeV. \quad (23)$$

In these expressions, $I = cE^2/4\pi$ is the laser intensity.

If $\beta_g < 1$ the front velocity is limited by the value $c\beta_g$ and for $B_E \to \infty$ the fast ion energy is finite according to Eqs. (15) and (20). The dashed line in Fig. 2 corresponds to the upper limit of accelerated ion energy which is equal to $2 m_\alpha c^2 \beta_g^2 / (1 - \beta_g^2)$.

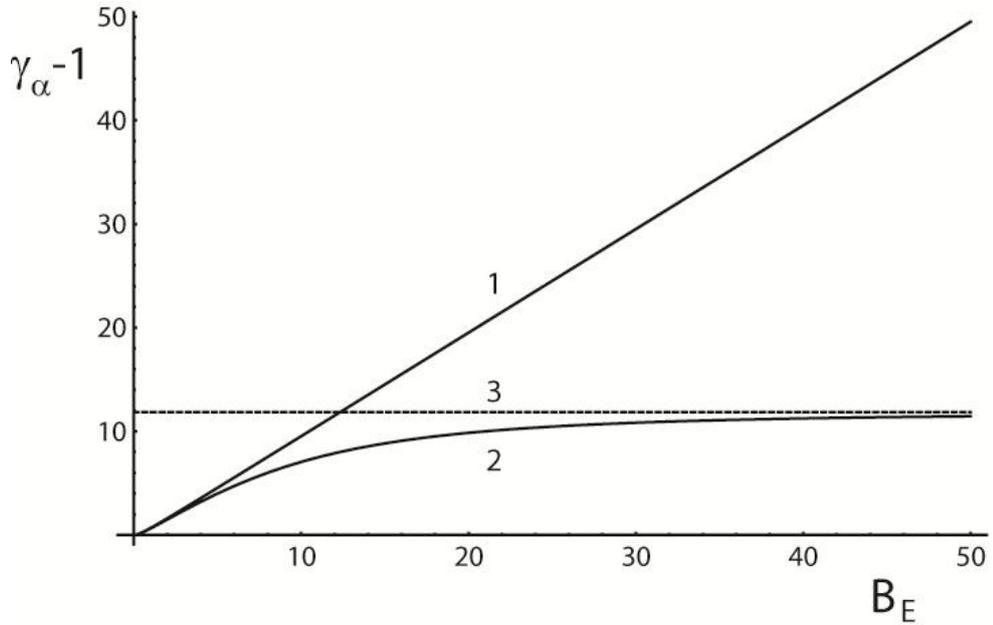

Fig. 2. Normalized ion kinetic energy, $\gamma_\alpha - 1$, vs the laser pulse amplitude, $B_E$: 1) for $\beta_g = 1$; 2) $\beta_g = 0.925$; 3) The dashed line corresponds to $2\beta_g^2 / (1 - \beta_g^2)$.

## 4. Instability of Counter-Propagating Ion Beams

*4a. Electrostatic Mode*

In the region ahead of the laser pulse front there is a multi-ion-stream configuration which is unstable against the electrostatic Buneman-like instability [32] (see discussion of various beam instabilities in Ref. [33]). This instability has been extensively studied with regard to a broad range of the problems related to laboratory and space plasmas [34] and, in particular, for explaining the nonlinear ion dynamics in plasmas irradiated by strong electromagnetic waves [35].

Linearizing the multi-stream hydrodynamic equations for an electron-two-ion-beam plasma we obtain the dispersion equation

$$1 - \frac{\omega_{pe}^2}{\bar{\omega}^2} - \frac{\omega_{p\alpha}^2}{2}\left[\frac{1}{\left(\bar{\omega} - \bar{k}_\parallel c\beta\right)^2} + \frac{1}{\left(\bar{\omega} + \bar{k}_\parallel c\beta\right)^2}\right] = 0, \qquad (24)$$

which gives the relationship between the frequency, $\bar{\omega}$, and wave number, $\bar{k}_\parallel$, of the perturbations in the boosted frame of reference. Here $\omega_{pe}^2$ and $\omega_{p\alpha}^2$ are equal to $4\pi n_0 e^2 \gamma / m_e$ and $4\pi n_0 Z_\alpha e^2 / m_\alpha \gamma^2$, respectively. The wave number, $\bar{k}_\parallel$, corresponds to perturbations propagating along the x-axis parallel to the ion beam velocity. We take into account that the electron and ion densities in the boosted frame of reference are larger by a factor $\gamma$ than those in the laboratory frame of reference. Here we have assumed an isotropization, fast on the ion time scale, of the electron component due to the two-stream instability (e.g. see [36]).

The roots of Eq. (24) can be expressed in the form

$$\bar{\omega} = \bar{\omega}' + i\bar{\omega}''. \qquad (25)$$

The dependence of the real and imaginary parts of the frequency, $\bar{\omega}' = \text{Re}[\bar{\omega}]$ and $\bar{\omega}'' = \text{Im}[\bar{\omega}]$, on the wave number is presented in Fig. 3.

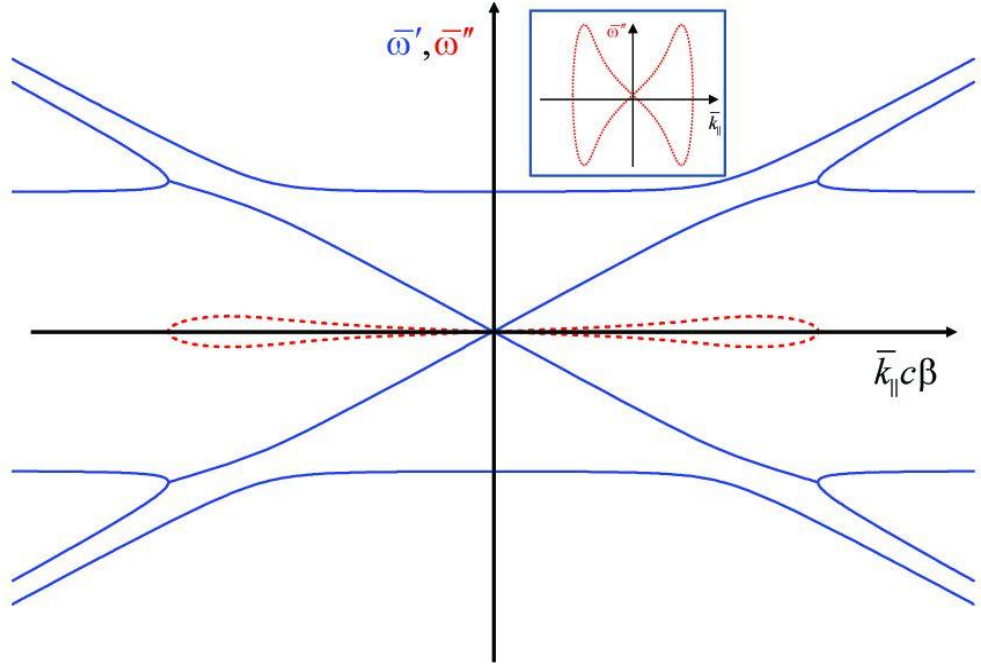

Fig. 3. (Colour online). Real (blue) and imaginary parts (dashed, red) of the frequency vs the wave number in the boosted frame of reference. Inset: Close-up of the frequency imaginary part, $\bar{\omega}''(\bar{k}_\parallel)$.

The multi-ion-stream configuration is unstable against the perturbations with wave number in the range $-\omega_{pe}/c\beta < \bar{k}_\parallel < \omega_{pe}/c\beta$. The growth rate has a maximum,

$$\max\{\mathrm{Im}[\bar{\omega}]\} \approx \left(\frac{\omega_{pe}\omega_{p\alpha}^2}{4}\right)^{1/3}, \tag{26}$$

at $\bar{k}_\parallel \approx \omega_{pe}/c\beta$ with the real part of the frequency approximately equal to $\mathrm{Re}[\bar{\omega}] \approx \bar{k}_\parallel c\beta$. In the long-wavelength (non-resonant) regime, at $\bar{k}_\parallel \to 0$ the growth of this instability is given by,

$$\mathrm{Im}[\bar{\omega}] \approx \left(\frac{\omega_{pe}}{\omega_{p\alpha}}\right)\bar{k}_\parallel c\beta. \tag{27}$$

The growth rate dependence on the wave number in the limits described by Eqs. (26) and (27) is seen in the inset in Fig. 3.

For long enough laser pulses with $\tau_{las} > 1/\max\{\mathrm{Im}[\omega]\}$, the instability development leads to the isotropization of the ion momentum distribution in the moving frame of reference, which results in formation in the laboratory frame of an ion beam with average energy equal to $m_\alpha c^2/(1-\beta^2)^{1/2}$ and beam energy width of the order of $m_\alpha c^2(1+\beta^2)/(1-\beta^2)$.

In the laboratory frame the instability growth rate and the exponential scale length can be obtained by using a Lorentz transformation. As a result we obtain for the space-time dependence of the perturbations

$$f \sim \exp\left\{\frac{\bar{\omega}''(t - x\beta/c) - i\left[\left(\bar{\omega}' + \bar{k}_{\|}\beta c\right)t - \left(\bar{k}_{\|} + \bar{\omega}'\beta/c\right)x\right]}{\sqrt{1-\beta^2}}\right\}. \qquad (28)$$

Above we analyzed the instability of two counter-penetrating ion beams considering the time development of initial perturbations. Time dependent perturbations can also be launched from the boundary causing the growth in space of an unstable mode. In the case under consideration the perturbations are generated by the laser field that forces ions and electrons to oscillate with the laser frequency at the vacuum-plasma interface. In order to find the spatial structure of the unstable mode we solve Eq. (24) with respect to the complex wave number,

$$\bar{k}_{\|} = \bar{k}_{\|}' + i\bar{k}_{\|}'', \qquad (29)$$

assuming the frequency, $\bar{\omega}$, to be fixed and real. Here we analyse the unstable modes in the frame of reference co-moving with the shock wave front. Then we obtain

$$\bar{k}_{\|} = \pm \frac{\bar{\omega}}{\beta c}\sqrt{1 + \frac{\omega_{p\alpha}^2 \pm \omega_{p\alpha}\sqrt{8\left(\bar{\omega}^2 - \omega_{pe}^2\right) + \omega_{p\alpha}^2}}{2\left(\bar{\omega}^2 - \omega_{pe}^2\right)}}. \qquad (30)$$

The dependence of the real and imaginary parts of the wave number, $\bar{k}_{\|}' = \text{Re}[\bar{k}_{\|}]$ and $\bar{k}_{\|}'' = \text{Im}[\bar{k}_{\|}]$, on the frequency, $\bar{\omega}$, is presented in Fig. 4.

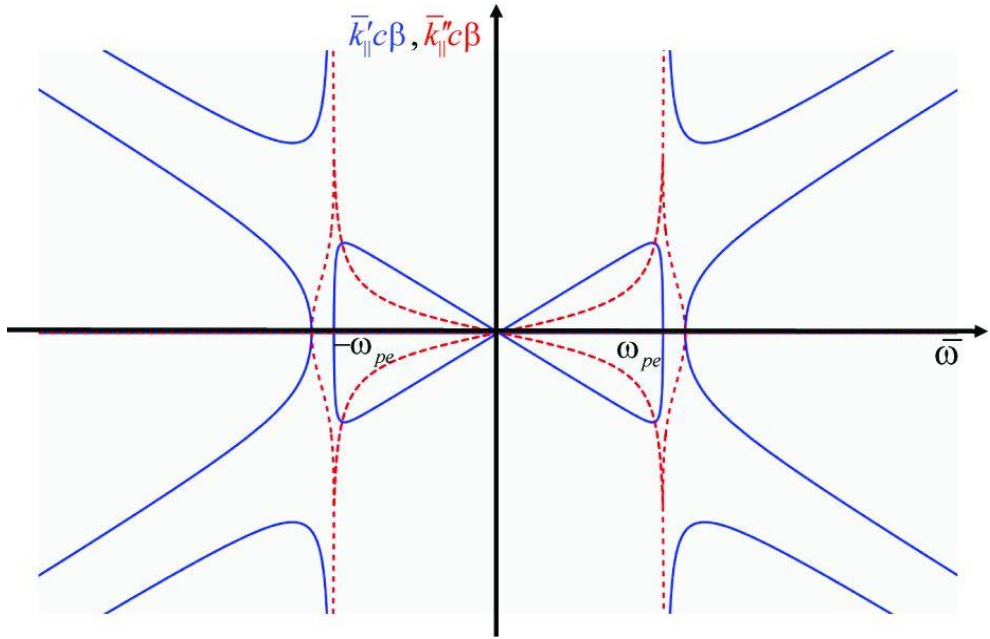

Fig. 4. (Colour online). Real (blue) and imaginary (red, dashed) parts of the wave number vs frequency in the boosted frame of reference.

As we see, in the high frequency limit, $\bar{\omega} \gg \omega_{pe}, \omega_{p\alpha}$, the wave number is real. Asymptotically, as $\bar{\omega} \to \infty$ we have for four branches of $\bar{k}(\bar{\omega})$

$$\bar{k}_\parallel = \pm \frac{\bar{\omega}}{\beta c}\left(1 \pm \frac{\omega_{p\alpha}}{2^{1/2}\bar{\omega}}\right). \tag{31}$$

The imaginary part of the wave number is not equal to zero for $\bar{\omega}^2 < \omega_{pe}^2 - \omega_{p\alpha}^2/8$. At the resonance, $\bar{\omega} \to \omega_{pe}$, the real and imaginary parts of the wave number tend either to infinity

$$\bar{k}'_\parallel \underset{\bar{\omega}\to\omega_{pe}+\varepsilon}{\approx} \pm\frac{\bar{\omega}}{\beta c}\frac{\omega_{p\alpha}}{\sqrt{\bar{\omega}^2-\omega_{pe}^2}}, \qquad k''_\parallel \underset{\omega\to\omega_{pe}-\varepsilon}{\approx} \pm\frac{\bar{\omega}}{\beta c}\frac{\omega_{p\alpha}}{\sqrt{\omega_{pe}^2-\bar{\omega}^2}}, \tag{32}$$

or to zero

$$\bar{k}'_\parallel \underset{\omega\to\omega_{pe}-\varepsilon}{\approx} \pm\frac{\bar{\omega}}{\beta c}\frac{8\left(\bar{\omega}^2-\omega_{pe}^2\right)}{\omega_{p\alpha}^2}. \tag{33}$$

In the low frequency limit, $\bar{\omega} \to 0$, assuming $\omega_{pe} \gg \omega_{p\alpha}$, we obtain

$$\bar{k}_\parallel \underset{\bar{\omega}\to 0}{\approx} \pm\frac{\bar{\omega}}{\beta c}\left(1 \pm i\frac{\omega_{p\alpha}}{\omega_{p\alpha}}\right). \tag{34}$$

This type of dependence of the imaginary and real parts of the wave number on the mode frequency corresponds, according to criteria formulated in Ref. [37], to a convective instability where waves with frequency $\omega < \omega_{pe}$ propagating from a source located at $x = 0$ are amplified in the direction $x > 0$, i.e. "down" the ion beam.

*4b. Electromagnetic Mode*

Counter- propagating beams of charged particles can also be unstable with respect to an electromagnetic type instability leading to the filamentation of the ion flows. This Weibel-type instability [38] in collisionless plasmas was extensively studied in regard to a broad range of problems of interest for space and laboratory plasmas (e.g. see Ref. [39] and literature quoted therein). This instability is due to the repulsion of oppositely directed electric currents. In the case under consideration, the oppositely directed electric currents are carried by the counter-propagating ion beams.

We take the perturbation dependence on the space and time coordinates in the co-moving with the shock front frame of reference to be of the form

$$f \sim \exp\left(-i\bar{\omega}t + ik_\perp y\right), \tag{35}$$

i.e. we assume that the perturbations are homogeneous in the direction of the ion beam motion. Following Ref. [32] we can easily find the dispersion equation for the frequency and wave number. It reads

$$\bar{\omega}^2 \left( \bar{\omega}^2 - k_\perp^2 c^2 - \omega_{pe}^2 - \omega_{p\alpha}^2 \right) - \left( \bar{\omega}^2 + k_\perp^2 c^2 \beta^2 \right) \omega_{p\alpha}^2 = 0, \qquad (36)$$

where $\omega_{pe}^2$ and $\omega_{p\alpha}^2$ are equal to $4\pi n_0 e^2 \gamma / m_e$ and $4\pi n_0 Z_\alpha e^2 / m_\alpha \gamma^2$, respectively. Its solution yields the real and the imaginary parts of the frequency

$$\bar{\omega}' + i\bar{\omega}'' = \pm \sqrt{ \frac{k_\perp^2 c^2 + \omega_{pe}^2 + \omega_{p\alpha}^2}{2} \pm \sqrt{ \frac{\left( k_\perp^2 c^2 + \omega_{pe}^2 + \omega_{p\alpha}^2 \right)^2}{4} + k_\perp^2 c^2 \beta^2 \omega_{p\alpha}^2 } } \ . \qquad (37)$$

The dependence of the real and imaginary parts of the frequency, $\bar{\omega}' = \mathrm{Re}[\bar{\omega}]$ and $\bar{\omega}'' = \mathrm{Im}[\bar{\omega}]$, on the wave number $k_\perp$ is presented in Fig. 5.

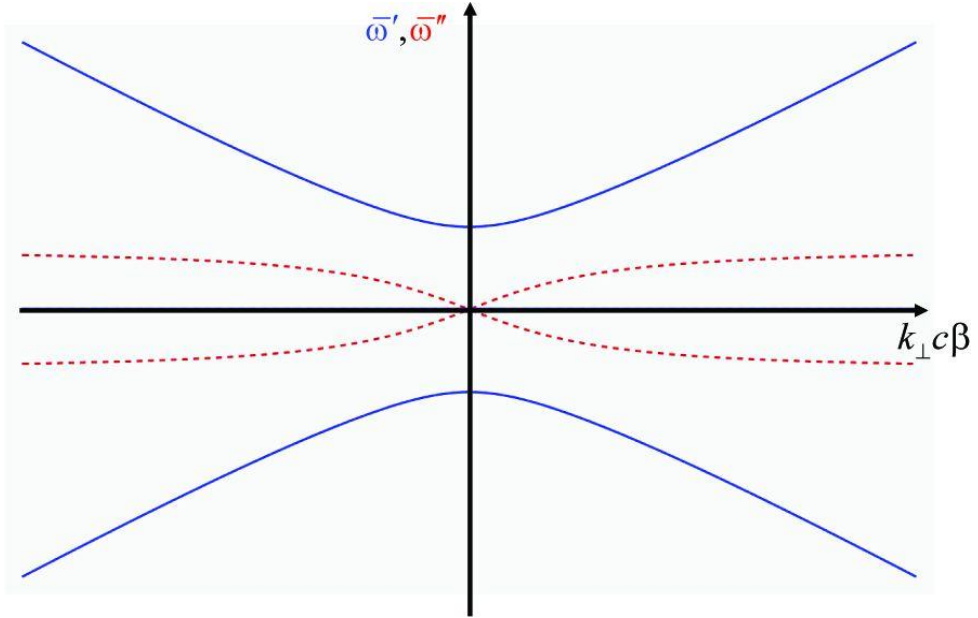

Fig. 5. (Colour online). Real (blue) and imaginary parts (dashed, red) of the frequency vs the wave number for $\sqrt{\omega_{pe}^2 + \omega_{p\alpha}^2} = 1$ and $\omega_{p\alpha} = 0.5$, in the boosted reference frame.

The characteristic scale length of the electromagnetic instability is equal to

$$k_\perp^{-1} \approx c\beta / \omega_{p\alpha}. \qquad (38)$$

In the long wavelength limit, when $k_\perp \to 0$, the growth rate and wave number are linearly proportional:

$$\mathrm{Im}[\bar{\omega}] \approx \left( \frac{\omega_{pe}}{\omega_{p\alpha}} \right) k_\perp c\beta. \qquad (39)$$

At $k_\perp \to \infty$ the growth rate tends to a constant equal to:

$$\max\{\text{Im}[\bar{\omega}]\} \approx \omega_{p\alpha}\beta. \tag{40}$$

In the laboratory frame of reference the instability growth rate and the exponentiation scale length can be obtained by using a Lorentz transformation. This gives for the coordinate-time dependence of the perturbations

$$f \sim \exp\left[\bar{\omega}''\left(\frac{t-\beta x/c}{\sqrt{1-\beta^2}}\right) - i\left(\bar{\omega}'\frac{t-\beta x/c}{\sqrt{1-\beta^2}} - k_\perp y\right)\right], \tag{41}$$

revealing an oblique pattern of the perturbations seen in the laboratory frame of references (see Fig. 7 c below).

*4b. Energy Spectrum of Fast Ions*

Due to the instability development the perturbation amplitude grows exponentially leading to the broadening of the energy spectrum of the fast ions. For the sake of simplicity, we assume that locally the fast ion distribution function can be approximated by a Gaussian function, i.e. we assume a weakly relativistic limit,

$$f_\alpha(p,x) = \frac{n_\alpha}{[4\pi m_\alpha T_\alpha(x)]^{1/2}} \exp\left[-\frac{(p-p_\alpha)^2}{2m_\alpha T_\alpha(x)}\right] \tag{42}$$

with $p_\alpha = 2m_\alpha c\beta/(1-\beta^2)$ and an effective temperature depending on the x coordinate as $T_\alpha(x) = T_0 \exp(x/L)$, where $T_0$ is determined by the amplitude of perturbations imposed by the shock wave front and, according to Eqs. (28) and (30), the scale length $L$ is equal to $L = \bar{k}_\parallel''/\sqrt{1-\beta^2}$. Integrating $f_\alpha(p,x)$ over the coordinate $x$ we obtain the ion energy spectrum,

$$N_\alpha(p) = \int_0^\infty f_\alpha(p,x)\,dx = \frac{n_\alpha L}{p-p_\alpha}\,\text{erf}\left[\frac{p-p_\alpha}{\sqrt{2m_\alpha T_0}}\right]. \tag{43}$$

Here $\text{erf}[z] = (2/\pi^{1/2})\int_0^z \exp(-t^2)\,dt$ is the error function [40]. In the vicinity of the maximum, $p - p_\alpha \to 0$, Eq. (43) gives

$$N_\alpha(p) \approx \frac{n_\alpha 2L}{(2\pi m_\alpha T_0)^{1/2}}\left[1 - \frac{(p-p_\alpha)^2}{6m_\alpha T_0}\right]. \tag{44}$$

At $(p-p_\alpha)^2/m_\alpha T_0 \gg 1$ the function $N_\alpha(p)$ decreases exponentially $N_\alpha(p) \propto \exp\left[-(p-p_\alpha)^2/2m_\alpha T_0\right]$. We see that the spectrum width is equal to the square root of the energy of the perturbations imposed by the shock wave front.

The ion energy spectrum is shown in Fig. 6 for $p_\alpha = 5$ and $T_0 = 0.25$. Note that similar energy spectra of the ions accelerated at the collisionless shock waves in laser plasmas have been seen in simulations and experiments [41].

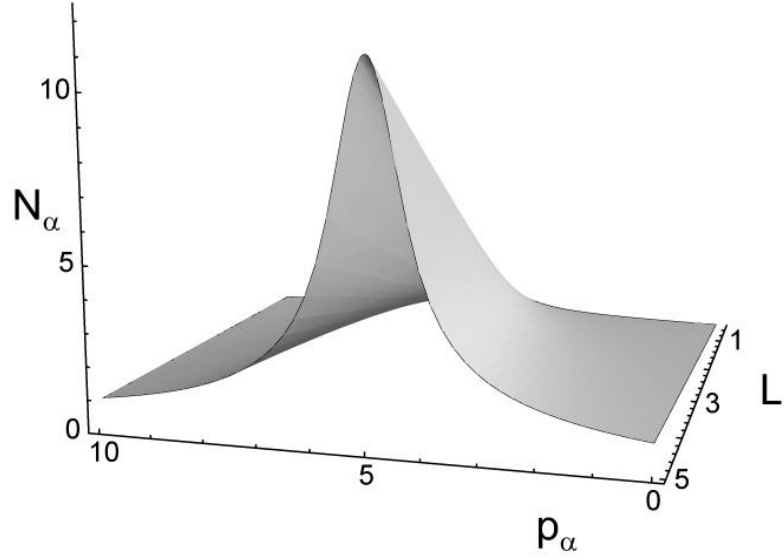

Fig. 6. Ion energy spectrum, $N_\alpha(p_\alpha, L)$ for $p_\alpha = 5$ and $2m_\alpha T_0 = 0.25$.

## 5. 2D PIC Simulation Results on the Hole Boring in an Overdense Plasma by Extremely High Power Laser Pulse

During its interaction with matter a super-high-power laser pulse, in addition to the instabilities caused by the fast charged particle beams, is subject to various other instabilities. Among them the most important is the relativistic self-focusing which results in the laser pulse channelling. In the context of the interaction with overdense targets it is also called "hole boring". It leads to the increase of the laser pulse amplitude and to the decrease of the electron density in the interaction region, which change the energy and the direction of the accelerated ions. A thorough study of these effects requires computer simulations.

We performed studies of the laser pulse interaction with high density targets using the two-dimensional (2D) particle-in-cell (PIC) code [42]. In these simulations, the laser pulse has normalized amplitude $a = 200$, pulse length $l_x = 100\lambda$ and width $l_y = 25\lambda$. It has a super Gaussian form and is circularly polarized. The plasma target density is equal to 256 $n_{cr}$. At $t = 0$ the plasma target is localized at $x > 10\lambda$. The simulation box dimensions are equal to $60\lambda \times 35\lambda$. The total number of quasi-particles is equal to $1.55 \times 10^7$. The plasma comprises electrons and protons with mass ratio equal to 1836. The simulation results for the parameters of interest are shown in Fig. 7, where we present the electromagnetic wave (Fig. 7 a,b) and the ion density (Fig. 7 c) distribution in the near axis region in the x,y plane at $t = 100(2\pi/\omega)$. We see that the ion density and the electromagnetic field at the plasma-vacuum interface are modulated in the transverse direction, i.e. along the y-axis, due to the development of an instability similar to that described in Refs. [10, 43]. The ion density is transversely modulated in the region of the counter-penetrating ion flows. In Fig. 7 b we see the generation of a small scale magnetic field which correlates with the ion density modulations. This can be attributed to the development of the the Weibel-like instability discussed above, which also predicts the oblique patterns in the ion density and magnetic field distribution (see Eq. (41)).

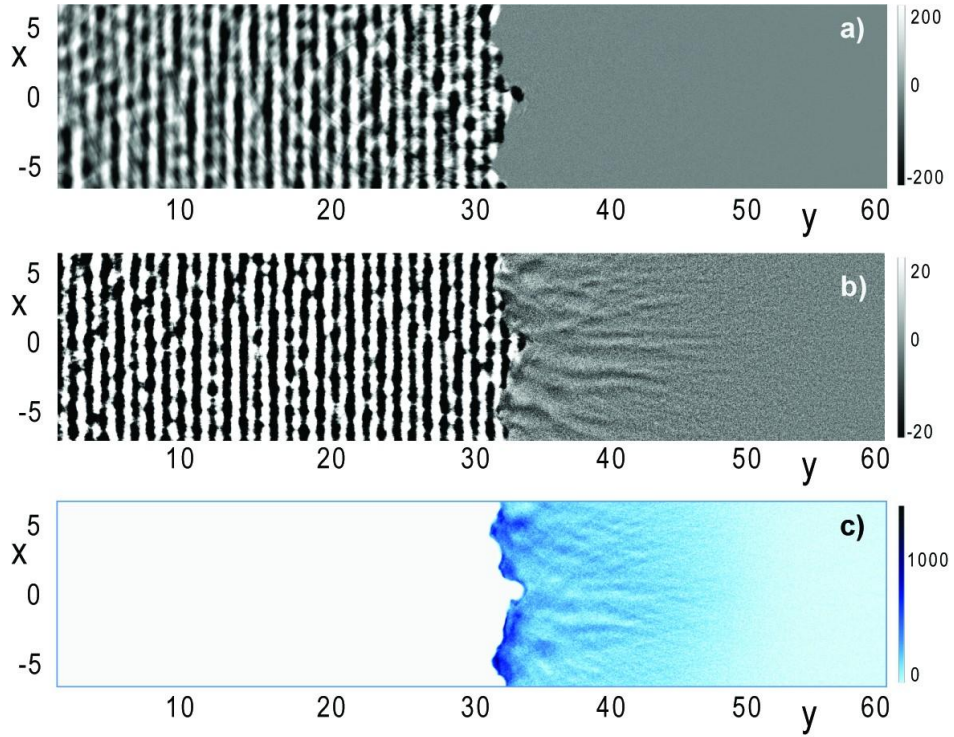

Fig. 7. (Colour online). Results of 2D PIC simulations. a) z-component of the electric field, b) z-component of the magnetic field, c) ion density distribution in the x,y plane at $t = 100(2\pi/\omega)$.

The region of the counter-penetrating ion flows corresponds to the collisionless shock wave. The ion energy spectrum shown in Fig. 8 is similar to the theoretically obtained spectrum plotted in Fig. 6. The number of accelerated ions is proportional to time and has a constant maximum energy and spectrum width. The maximum energy is about 180 MeV with a spectrum width of the order of 100 MeV. The phase plots in Fig. 9 e-h show distinctly the fast ion distribution which can be interpreted as caused by the development of the two-ion-beam instability discussed above. The transverse modulations of the fast ion density (Fig. 8 c) and of the transverse and longitudinal momentum (Fig. 9 a, b) seen in the region ahead of the shock wave front can be interpreted as caused by the electromagnetic instability as well as by the front corrugations due to the laser pulse filamentation.

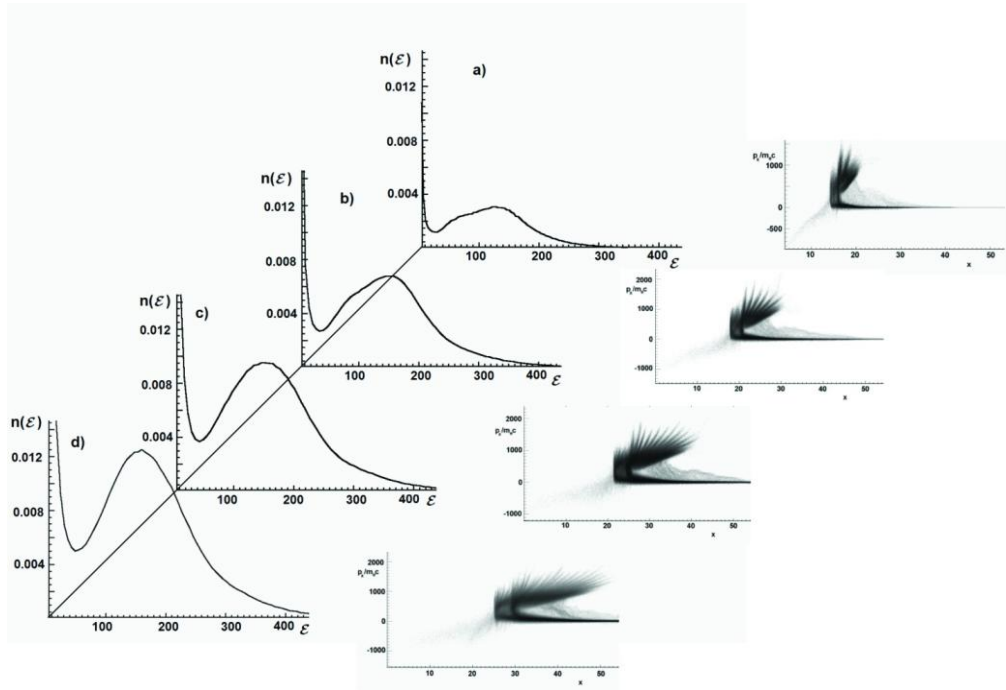

Fig. 8. The ion energy spectrum and $(x, p_x)$ phase plane at a, e) $t = 43.5(2\pi/\omega)$; b, f) $t = 62.5(2\pi/\omega)$; c, g) $t = 80(2\pi/\omega)$; d,h) $t = 100(2\pi/\omega)$.

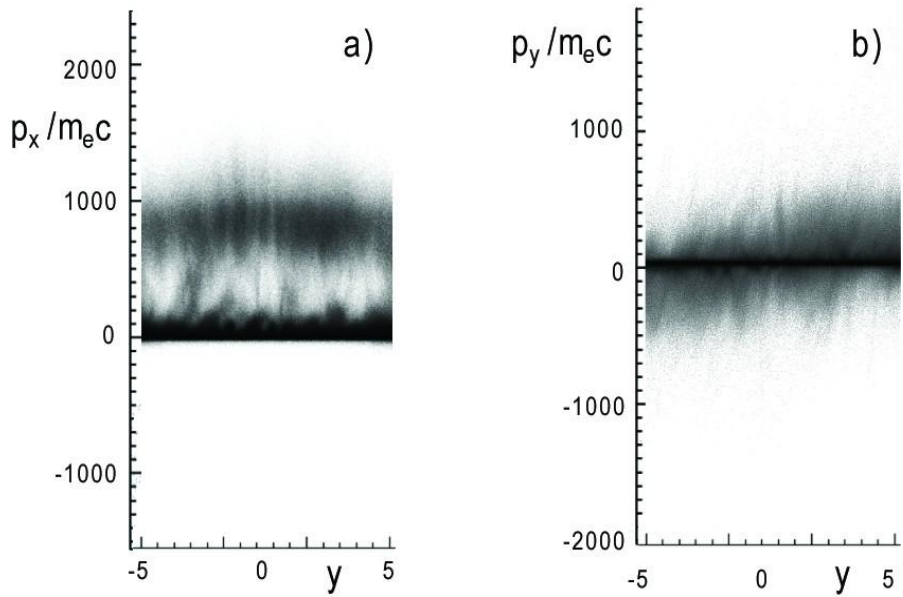

Fig. 9. The ion phase plane a) $(y, p_x)$ and b) $(y, p_y)$ at a, e) $t = 100(2\pi/\omega)$.

## 6. Conclusion

We presented theoretical models describing the RPDA regime of ion acceleration by slow super-intense laser pulses interacting with thin plasma slabs and with extended plasma targets. It is shown that the energy of the accelerated ions cannot exceed the energy corresponding to the electromagnetic wave group velocity. For example, when the electromagnetic wave propagates inside a waveguide of radius $r$, the gamma-factor corresponding to the wave group velocity is equal to, $1/\sqrt{1-\beta_g^2} = \omega r/1.84c$, which yields the value $\mathcal{E}_\alpha = 2m_\alpha c^2 \pi r/1.84\lambda$ for the limit on the fast ion energy. When the laser pulse interacts with an extended plasma target, the ions reflected from the laser pulse front form high energy beams. Such configuration is unstable with respect to electrostatic and electromagnetic modes, which results in the broadening of the ion energy spectrum and in the ion beam filamentation.

Further studies of the laser plasma interaction in the RPDA regime will contribute to the development of the new discipline of laboratory astrophysics [44], to high energy physics [3, 45], to thermonuclear fusion with laser accelerated ions, [46] to the development of the laser ion accelerators for hadron therapy [47], to the development of the coherent x-ray sources [48], to the study of the light sail mechanism for spacecraft propulsion [7, 8] and to investigation of the feasibility of using a laser radiation for preventing orbital debris-debris collisions [49].

**Acknowledgment**

The authors are grateful for fruitful discussions to J. Koga, G. Korn, A. Macchi, T. Nakamura, N. N. Rosanov, and S. G. Rykovanov. We appreciate support from the NSF under Grant No. PHY-0935197 and the US DOE under Contract No. DE-AC02-05CH11231.